\documentclass{article}

\usepackage{arxiv}

\usepackage{graphicx}%
\usepackage{multirow}%
\usepackage{amsmath,amssymb,amsfonts}%
\usepackage{amsthm}%
\usepackage{mathrsfs}%
\usepackage{xcolor}%
\usepackage{textcomp}%
\usepackage{manyfoot}%
\usepackage{booktabs}%
\usepackage{algorithm}%
\usepackage{algorithmicx}%
\usepackage{algpseudocode}%
\usepackage{listings}%
\usepackage{hyperref}
\usepackage[numbers]{natbib}
\usepackage{hyperref} 
\hypersetup{colorlinks = true, citecolor = blue}

\usepackage{siunitx}
\title{Is TabPFN the Silver Bullet for Insurance Pricing?}
\author{
	Bruno Deprez\thanks{Correspondening author: \href{mailto:bruno.deprez@kuleuven.be}{\texttt{bruno.deprez@kuleuven.be}}} \\
	KU Leuven\\
	University of Antwerp-imec
	\And
	Wouter Verbeke\\
	KU Leuven
	\And
	Tim Verdonck\\
	University of Antwerp-imec\\
	KU Leuven 
}

\begin{document}
	\maketitle
	
	\begin{abstract}
		Modelling claim frequency and severity for non-life insurance pricing predominantly relies on generalised linear models (GLMs), with gradient-boosted machines as the leading machine learning alternative. 
		Tabular foundation models (TFMs) present a fundamentally different inference paradigm. 
		By pre-training on large collections of synthetic datasets, TFMs enable inference on new data through in-context learning, without any dataset-specific preprocessing, fitting, or hyperparameter tuning. 
		This paper presents a first empirical evaluation of TabPFN for motor insurance pricing, benchmarking two versions of TabPFN against GLM and XGBoost on two publicly available motor third-party liability (MTPL) datasets. 
		Our results show that TabPFN does not consistently outperform established baselines: it falls behind the GLM on claim frequency, where its performance depends on the size of the in-context training set, but it is competitive on claim severity, where data are scarcer. Its inference times are substantially longer. Although tabular foundation models are promising, particularly in data-scarce settings, TabPFN is not yet ready to replace established actuarial methods for claim frequency.
	\end{abstract}
	
	
	\section{Introduction}
	Determining the technical price $\pi$ of a non-life insurance product requires an accurate estimation of the expected loss per unit of exposure, where $L$ is the total loss and $e$ the exposure. 
	The expected loss per unit of exposure is classically decomposed into claim frequency $F$ (the number of claims $N$ per unit of exposure) and claim severity $S$ (the average loss per claim):
	\begin{equation}\pi = \mathbb{E}\left(\frac{L}{e}\right) = \mathbb{E}\left( \frac{N}{e} \right)\cdot\mathbb{E}\left( \frac{L}{N} \mid N>0\right) = \mathbb{E}\left( F \right) \cdot \mathbb{E}\left( S \right).\end{equation}
	
	Frequency and severity are typically modelled with generalised linear models (GLMs) or generalised additive models (GAMs) in practice~\citep{de2008generalized, Holvoet03072025}. Both are highly transparent, given that their outputs are easily explained to stakeholders based on the estimated model parameters~\citep{Hainaut26112022}. 
	
	Tree-based learners, such as decision trees and random forests, capture non-linearity and interactions automatically while retaining transparency to some extent~\citep{Trufin2,chevalierboosting}, and can be used to construct a competitive GLM~\citep{Henckaerts03042021}. 
	Deep learning has been adopted more slowly. 
	On tabular data including around 10k samples, tree-based methods outperform it~\citep{Grinsztajn_tree_vs_dl}, and standard neural networks yield overly smooth predictions~\citep{Holvoet03072025}. 
	Research has integrated actuarial knowledge, as in combined actuarial neural networks~\citep{schelldorfer2019nesting}, the LocalGLMnet~\citep{Richman02012023} or the credibility transformer~\citep{credibilityTransformer, padayachy2026incontextlearningenhancedcredibility}. 	
	A shared limitation of these methods is the need for dataset-specific preprocessing, fitting, and hyperparameter tuning.
	
	A fundamentally different inference paradigm has recently emerged with tabular foundation models (TFMs)~\citep{pmlr-v267-qu25d, tabpfn}.
	Based on Prior-data Fitted Networks (PFNs), a class of transformer-based models pre-trained on multiple datasets, these models allow for in-context learning~\citep{muller2022transformers}.
	A promising TFM is TabPFN~\citep{hollmann2023tabpfn, tabpfn}, which is pre-trained on millions of synthetic tabular datasets and requires no dataset-specific fitting (Section~\ref{sec:tabpfn}).
	
	\citet{tabpfn} illustrate that TabPFN with default hyperparameters on average outperforms tuned state-of-the-art tabular models.
	However, these benchmarks only evaluate prediction accuracy and thus provide no evidence on its suitability for insurance pricing.
	
	One risk of using TabPFN out of the box is that insurance pricing is reduced to a generic regression problem. 
	TabPFN does not yet incorporate all aspects of exposure, sparse claim counts, heterogeneous conditional variance, and distribution-specific loss functions, which are crucial for frequency and severity pricing. 
	
	To our knowledge, in the peer-reviewed literature, only \citet{10475046} have applied TabPFN in insurance, framing cross-selling of health insurance as a binary classification task. 
	Whether TabPFN is a competitive alternative for the core actuarial problem of pricing remains an open question. 
	This paper presents a first empirical assessment of TabPFN on non-life insurance pricing data, benchmarking two versions of it against GLM and extreme gradient boosting (XGBoost)~\citep{chen2016xgboost} on two publicly available motor third-party liability (MTPL) datasets.
	
	\section{TabPFN}\label{sec:tabpfn}
	TabPFN approximates Bayesian inference~\citep{muller2022transformers, tabpfn}.
	Given a prior $p(\varphi)$ over data-generating mechanisms $\varphi$ and a training context $\mathcal{D} = (\boldsymbol{X}_{\mathrm{train}}, \boldsymbol{y}_{\mathrm{train}})$, the prediction for new data $\boldsymbol{X}_{\mathrm{test}}$ is the posterior predictive distribution (PPD)
	\begin{equation}
		p(\hat{\boldsymbol{y}}\mid \boldsymbol{X}_{\mathrm{test}}, \mathcal{D}) = \int p(\hat{\boldsymbol{y}}\mid \boldsymbol{X}_{\mathrm{test}}, \varphi)\, p(\varphi \mid \mathcal{D})\, \mathrm{d}\varphi.
	\end{equation}
	This differs from Bayesian inference in actuarial science, such as credibility theory and Bayesian GLMs~\citep{denuit2019effective}, where a prior is placed on the parameters of a fixed parametric model. 
	TabPFN instead places a prior on entire data-generating mechanisms, and the resulting posterior predictive distribution is approximated by a neural network rather than derived analytically or by sampling.
	Hence, TabPFN has no explicit likelihood and no posterior distribution over interpretable parameters.
	TabPFN learns to approximate the PPD once, during pre-training on millions of synthetic datasets generated from random structural causal models, by predicting held-out rows of each dataset given the other rows as context~\citep{tabpfn,grinsztajn2026tabpfn3technicalreport}.
	At prediction time, the weights are fixed and the training set is passed through the network together with the test rows.
	For regression, TabPFN outputs a discretised predictive distribution without parametric assumptions~\citep{tabpfn}, of which we use the mean, averaged by default over an ensemble of forward passes with different preprocessing and feature orders.
	
	We consider two versions of TabPFN.
	TabPFN-v2.6 attends across the features of each row and across the training rows of each feature~\citep{tabpfn}, and supports up to 100,000 samples and 2,000 features, so larger training sets must be subsampled.
	TabPFN-v3 instead compresses each sample into a single row embedding before in-context learning, trading feature capacity for sample capacity~\citep{grinsztajn2026tabpfn3technicalreport}.
	It scales to 1,000,000 samples provided that the feature count remains below 200.
	
	Both versions assume that the data resemble the synthetic prior, which is not tailored to insurance data, and that observations are exchangeable and equally weighted. 
	Moreover, TabPFN supports neither offsets nor observation weights.
	In addition, the context size is limited, and the inference time grows with the context size, since the training set is passed through the network again for every prediction.
	
	A further limitation concerns licensing: the TabPFN code is released under the Apache 2.0 licence, but the weights of v2.6 and v3 fall under Prior Labs' non-commercial licences, which at the time of writing permit research and evaluation, as in this study, but not commercial or production use.
	Insurers that want to deploy TabPFN for pricing therefore need a commercial licence from Prior Labs.
	
	\section{Experimental Set-up}
	We benchmark TabPFN-v2.6 and TabPFN-v3 against a Poisson/gamma GLM and XGBoost on the French freMTPL2 and Belgian beMTPL97 datasets from the CASdatasets package~\citep{CAS}. These are public MTPL benchmarks that have been used in several actuarial studies~\citep{Richman02012023, Holvoet03072025,Henckaerts03042021,HENCKAERTS2022117230,credibilityTransformer,Lindholm_Palmquist_2024,chevalierboosting,brauertransformer}, and they differ in size (about 678,000 and 163,000 policies), which allows us to examine how performance and inference time depend on the amount of training data. 
	Comparing both versions shows whether the newer architecture and its extended context size improve predictive performance and computational efficiency.
	
	TabPFN is designed to operate on raw inputs, so preprocessing is applied to GLM and XGBoost only.
	We apply ordinal encoding for ordered categoricals, one-hot encoding for nominal categoricals, and no transformation of numerical features. XGBoost uses the fixed hyperparameters of \citet{Henckaerts03042021} (maximum depth 3, 500 trees, learning rate 0.01, subsample ratio 0.75), without dataset-specific tuning. TabPFN is used with the default settings of each version, again without tuning. 
	
	The models differ in how they account for exposure.
	GLM and XGBoost incorporate the claim count $N_i$ with $\ln e_i$ as an offset, so that the expected number of claims is proportional to the exposure~\citep{wuthrich2025data, Henckaerts03042021}.
	TabPFN, however, supports neither an offset nor observation weights, so exposure cannot enter the model in the same way.
	We include the exposure $e_i$ as an input feature and take the claim count $N_i$ as the target~\citep{priorlabs_insurance}.
	We do not model the claim rate $N_i/e_i$ directly, since dividing by small exposures produces extreme values.
	The claim rate is obtained by evaluating every policy at a full year of exposure, which places the predictions on the same annualised scale as those of the GLM and XGBoost.
	Note that this is not equivalent to an offset: whereas the GLM and XGBoost impose proportionality between exposure and the expected claim count, TabPFN learns the effect of exposure from the data.
	Another option would be to limit the data to observations with exposure equal to 1, but we did not explore this further since it is not in line with how the models would be used in reality. 
	
	For severity, the GLM and XGBoost are weighted by the number of claims $N_i$, whereas TabPFN is trained unweighted, as it does not support observation weights.
	To account for the right-skewed claim amounts, TabPFN is fitted on $\ln(1 + y_i)$ using its built-in target transformation~\citep{priorlabs_insurance}.
	TabPFN maps its full predictive distribution back to the original scale, and we take its mean as the predicted severity.
	
	All models are trained and evaluated using 5-fold cross-validation, with performance summarised through the root mean squared error (RMSE), distributional deviance, and computational cost.
	The RMSE is a purely predictive loss. 
	For frequency, it is computed on the claim rates $y_i = N_i / e_i$ and weighted by the exposures $e_i$; for severity, it is computed on the average severities $y_i$ without weights.
	
	The deviance, in contrast, assesses how well the predictions fit the distributional assumptions classically adopted for frequency and severity modelling~\citep{wuthrich2025data, Henckaerts03042021}. 
	Claim frequency is typically assumed to follow a Poisson distribution~\citep{wuthrich2025data, Henckaerts03042021, Holvoet03072025}, so we use exposure-weighted Poisson deviance for evaluation: 
	\begin{equation}
		D_\text{Poisson}(f(\boldsymbol{x}), y) = \frac{2}{n_f}\sum_{i=1}^{n_f} \left( y_i\ln\frac{y_i}{f(\boldsymbol{x}_i)}-\left(y_i- f(\boldsymbol{x}_i)\right) \right).
	\end{equation}
	Claim severity is typically assumed to follow a gamma distribution~\citep{wuthrich2025data, Henckaerts03042021, Holvoet03072025}, so we use gamma deviance for evaluation:
	\begin{equation}
		D_\text{gamma}(f(\boldsymbol{x}), y) = \frac{2}{n_s}\sum_{i=1}^{n_s} N_i \left( \frac{y_i-f(\boldsymbol{x}_i)}{f(\boldsymbol{x}_i)} - \ln\frac{y_i}{f(\boldsymbol{x}_i)} \right), 
	\end{equation}
	where $N_i$ is the number of claims for policyholder $i$. 
	Here, $f(\boldsymbol{x}_i)$ is the prediction for policyholder $i$, and $n_f$ and $n_s$ are the numbers of policies, respectively policies with at least one claim, in the evaluation fold.
	Other distributional assumptions exist, but a full comparison falls outside the scope of this paper.
	
	Insurance pricing models operate under latency constraints, as a quote requested online must be returned within seconds, making inference time a critical operational metric. 
	We therefore analyse computation time in addition to the above performance metrics.
	For GLM and XGBoost we combine training and inference time.
	TabPFN reports only inference time given that no training happens for in-context learning. 
	
	To analyse the role of the context size, we vary it for the frequency data, where each training fold contains approximately 542,400 (freMTPL2) and 130,600 (beMTPL97) policies. 
	We evaluate context sizes of 2,000, 5,000, 10,000, 50,000 and 100,000, the latter being the maximum context size of TabPFN-v2.6, and additionally use the full training fold for TabPFN-v3. 
	The severity data are much smaller, with 24,944 (freMTPL2) and 18,276 (beMTPL97) policies with at least one claim, so each training fold (approximately 20,000 and 14,600 policies) lies well within the maximum context of both versions and is used in full.
	
	All experiments are implemented in Python and run on an Intel Xeon Platinum 8360Y CPU and NVIDIA A100 SXM4 GPU of the Flemish Supercomputer Center. 
	The implementation is available on GitHub (\url{https://github.com/B-Deprez/tabpfn_insurance}).
	
	\section{Results}
\begin{figure}
		\centering
		\includegraphics[width=0.95\linewidth]{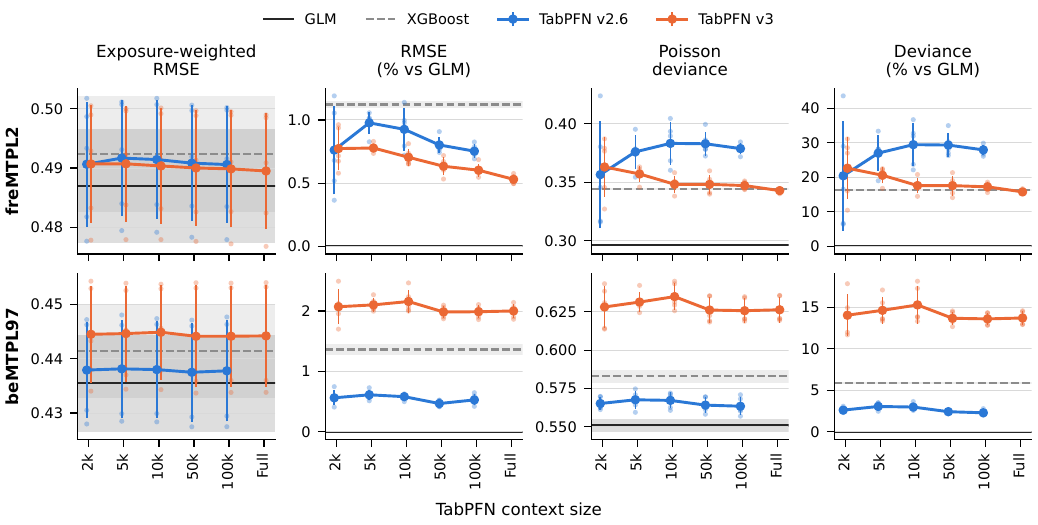}
		\caption{Frequency: results by context size, absolute and as fold-wise \% difference to the GLM. Dots are individual folds.}
		\label{fig:frequency}
	\end{figure}

	The frequency results are presented in Table~\ref{tab:results_freq} and Figure~\ref{fig:frequency}. 
	TabPFN-v2.6 attains lower RMSE than XGBoost on the Belgian dataset, but does not achieve the performance of the GLM.
	TabPFN-v3 matches or slightly improves on the RMSE of XGBoost on the French data at every context size, but performs worse than the other models on the Belgian data. 
	Across TabPFN configurations, context size does not seem to affect RMSE values much. 
	
	\begin{table}
		\centering
		\caption{Frequency results. TabPFN uses $n$ policies of the training fold as context; GLM and XGBoost use the full fold. \textbf{Bold}: best per column.}
		\label{tab:results_freq}
		\begin{tabular}{lcccc}
			\toprule
			& \multicolumn{2}{c}{freMTPL2} & \multicolumn{2}{c}{beMTPL97} \\
			\cmidrule(lr){2-3} \cmidrule(lr){4-5}
			Model & RMSE & Pois.\ Dev. & RMSE & Pois.\ Dev. \\
			\midrule
			GLM & $\mathbf{0.487 \pm 0.010}$ & $\mathbf{0.296 \pm 0.001}$ & $\mathbf{0.435 \pm 0.009}$ & $\mathbf{0.551 \pm 0.004}$ \\
			XGBoost & $0.492 \pm 0.010$ & $0.344 \pm 0.001$ & $0.441 \pm 0.009$ & $0.583 \pm 0.004$ \\
			\addlinespace[4pt]
			\multicolumn{5}{l}{\textit{TabPFN-v2.6} (context size)} \\
			\quad 2,000 & $0.491 \pm 0.011$ & $0.356 \pm 0.046$ & $0.438 \pm 0.009$ & $0.565 \pm 0.004$ \\
			\quad 5,000 & $0.492 \pm 0.010$ & $0.376 \pm 0.015$ & $0.438 \pm 0.009$ & $0.568 \pm 0.006$ \\
			\quad 10,000 & $0.491 \pm 0.010$ & $0.383 \pm 0.018$ & $0.438 \pm 0.009$ & $0.567 \pm 0.005$ \\
			\quad 50,000 & $0.491 \pm 0.010$ & $0.383 \pm 0.011$ & $0.437 \pm 0.009$ & $0.564 \pm 0.005$ \\
			\quad 100,000 & $0.491 \pm 0.010$ & $0.379 \pm 0.005$ & $0.438 \pm 0.009$ & $0.563 \pm 0.006$ \\
			\addlinespace[4pt]
			\multicolumn{5}{l}{\textit{TabPFN-v3} (context size)} \\
			\quad 2,000 & $0.491 \pm 0.010$ & $0.363 \pm 0.026$ & $0.444 \pm 0.009$ & $0.628 \pm 0.014$ \\
			\quad 5,000 & $0.491 \pm 0.010$ & $0.357 \pm 0.007$ & $0.445 \pm 0.009$ & $0.631 \pm 0.007$ \\
			\quad 10,000 & $0.490 \pm 0.010$ & $0.348 \pm 0.007$ & $0.445 \pm 0.009$ & $0.635 \pm 0.009$ \\
			\quad 50,000 & $0.490 \pm 0.010$ & $0.348 \pm 0.008$ & $0.444 \pm 0.009$ & $0.626 \pm 0.009$ \\
			\quad 100,000 & $0.490 \pm 0.010$ & $0.347 \pm 0.003$ & $0.444 \pm 0.009$ & $0.626 \pm 0.008$ \\
			\quad Full fold & $0.490 \pm 0.010$ & $0.343 \pm 0.002$ & $0.444 \pm 0.009$ & $0.626 \pm 0.008$ \\
			\bottomrule
		\end{tabular}
	\end{table}

	Also in terms of Poisson deviance GLM performs best on both datasets.
	XGBoost outperforms all TabPFN-v2.6 configurations and outperforms TabPFN-v3 at every context size except the full training fold on the French dataset, while TabPFN-v2.6 outperforms XGBoost at every context size on the Belgian dataset.
	The deviance of TabPFN is more sensitive to the context size than its RMSE. 
	For TabPFN-v3 on the French data, the deviance improves almost steadily with the context size, from 0.363 at 2,000 policies to 0.343 on the full training fold.
	For TabPFN-v2.6, in contrast, it follows a non-monotonic pattern: performance worsens from 2,000 to 10,000 policies and only improves slightly at 100,000. 
	Similar non-monotonic patterns were observed by \citet{baesens2026foundationmodelscreditrisk} for TabPFN in their benchmark on credit risk modelling. 
	This likely reflects distributional mismatch between small subsamples and the full training fold, and warrants further research into the underlying cause.
	
	\begin{table}
		\centering
		\caption{Severity results. TabPFN uses the full training fold as context. \textbf{Bold}: best per column.}
		\label{tab:results_sev}
		\begin{tabular}{lcccc}
			\toprule
			& \multicolumn{2}{c}{freMTPL2} & \multicolumn{2}{c}{beMTPL97} \\
			\cmidrule(lr){2-3} \cmidrule(lr){4-5}
			Model & RMSE & Gam.\ Dev. & RMSE & Gam.\ Dev. \\
			\midrule
			GLM & $21{,}707 \pm 21{,}556$ & $1.755 \pm 0.263$ & $3{,}479 \pm 199$ & $2.046 \pm 0.067$ \\
			XGBoost & $\mathbf{21{,}670 \pm 21{,}531}$ & $\mathbf{1.598 \pm 0.141}$ & $3{,}480 \pm 200$ & $2.055 \pm 0.079$ \\
			\addlinespace[2pt]
			TabPFN-v2.6 & $21{,}679 \pm 21{,}567$ & $1.654 \pm 0.229$ & $3{,}501 \pm 189$ & $2.092 \pm 0.043$ \\
			TabPFN-v3 & $21{,}686 \pm 21{,}572$ & $1.737 \pm 0.306$ & $\mathbf{3{,}478 \pm 199}$ & $\mathbf{2.041 \pm 0.064}$ \\
			\bottomrule
		\end{tabular}
	\end{table}

	\begin{figure}
		\centering
		\includegraphics[width=0.95\linewidth]{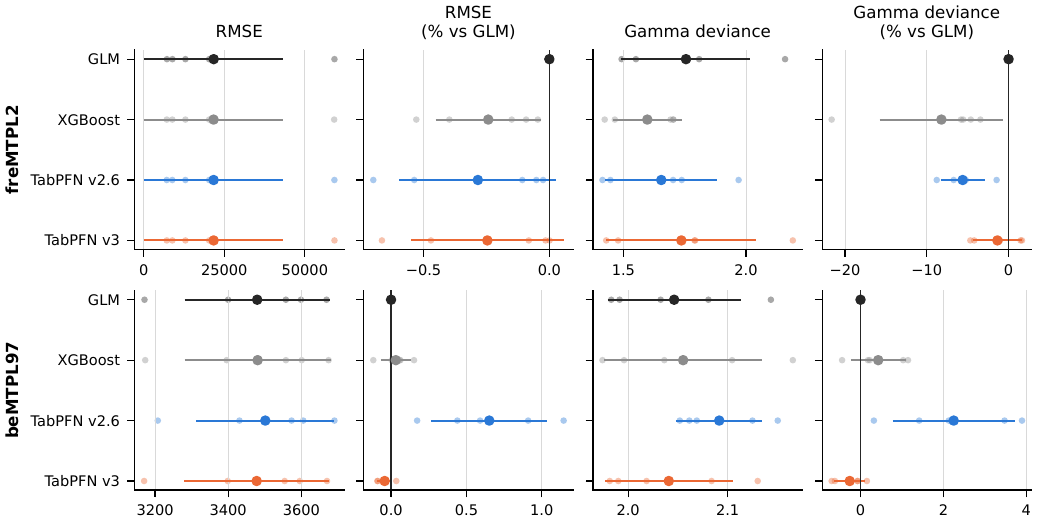}
		\caption{Severity: results, absolute and as fold-wise \% difference to the GLM. Dots are individual folds.}
		\label{fig:severity}
	\end{figure}
	
	The results on the severity data paint a different picture (Table~\ref{tab:results_sev} and Figure~\ref{fig:severity}).
	The GLM model performs the worst of all on the French dataset, while the XGBoost model performs best, followed by TabPFN-v2.6. 
	On the Belgian dataset, TabPFN-v3 performs best while TabPFN-v2.6 performs worst. 
	
	The standard deviations reported in Table~\ref{tab:results_sev} show strong variance in performance across folds, especially on the French dataset. 
	Figure~\ref{fig:severity} illustrates that this is mostly influenced by one outlier.
	To mitigate this, we also look at the relative performance compared to the GLM model. 
	
	Relative to the GLM, the results on the French dataset partly favour TabPFN-v2.6.
	It seems that TabPFN-v2.6 has a better average performance than XGBoost in terms of RMSE, although the performance is still more spread out, with some folds on which TabPFN outperforms XGBoost and some on which it does not.
	The relative gamma deviance on average still favours XGBoost, but we see that this is due to one outlier.
	The results for TabPFN-v2.6 are much more stable, and on most folds it outperforms XGBoost. 
	
	The conclusions on the Belgian dataset do not change much when going to relative performance. 
	TabPFN-v3 is still the best model over most folds.
	
	\begin{figure}
		\centering
		\includegraphics[width=0.95\linewidth]{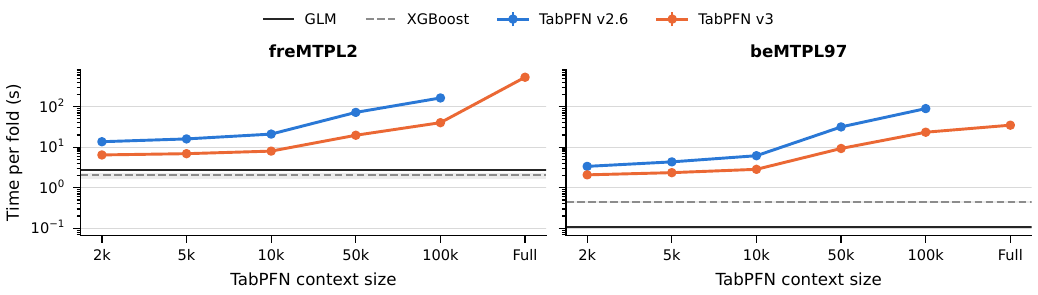}
		\caption{Training and inference time per fold (frequency, log scale).}
		\label{fig:time}
	\end{figure}
	
	Practical concerns extend beyond predictive power. 
	Inference time of TabPFN grows steeply with context size (Figure~\ref{fig:time}), which may limit its deployment in high-throughput, latency-sensitive environments. 
	Even at small context sizes, TabPFN's inference cost substantially exceeds the combined training and inference time of GLM and XGBoost, reaching several minutes at the model's capacity limit.
	Although these times cover scoring an entire fold rather than a single quote, the gap indicates that currently only GLM and XGBoost are suitable for real-time pricing applications at scale.
	We observe that the inference time is much lower for TabPFN-v3 compared to TabPFN-v2.6.
	
	\section{Conclusion}\label{sec:conclusion}
	We present a first empirical evaluation of TabPFN for non-life insurance pricing on two MTPL datasets.
	TabPFN does not consistently outperform established baselines. 
	Although TabPFN is competitive on the smaller severity datasets, it falls behind on frequency where GLM still performs best.
	TabPFN supports neither an offset nor observation weights, so exposure enters as a feature and its effect must be learned.
	In practice, TabPFN thus offers a trade-off.
	It requires no dataset-specific preprocessing, fitting, or hyperparameter tuning, returns a full predictive distribution, and is competitive when data are scarce, as for severity.
	These advantages come at the cost of long inference times on GPU hardware, a non-commercial licence for its weights, the lack of support for offsets and observation weights, and the absence of transparent coefficients.
	Therefore, TabPFN cannot yet replace established actuarial methods for claim frequency.
	
	Future work should investigate fine-tuning foundation models on insurance data using actuarial expert knowledge, extending the benchmark to additional datasets and TFMs, comparing with further methods proposed for these datasets, examining the downstream impact on premium setting and profitability, and analysing whether TabPFN recovers the same risk drivers as GLMs or yields complementary insights.
	
	Another important direction lies in developing tabular foundation models that handle exposure, sparse claim counts, heterogeneous conditional variance, and distribution-specific loss functions, which are crucial for insurance pricing. 
	
	Although our results do not favour TabPFN for claim frequency pricing, foundation models may prove more useful elsewhere in the insurance value chain, such as fraud detection using classification or anomaly detection, loss reserving with run-off triangles, or mortality projections in life insurance. 
	
	\section*{Acknowledgements}
	This work was supported by the Research Foundation – Flanders (FWO) [grant numbers 1SHEN24N and G015020N]. 
	The resources and services used in this work were provided by the VSC (Flemish Supercomputer Center), funded by the Research Foundation – Flanders (FWO) and the Flemish Government.

	\bibliographystyle{agsm}
	\bibliography{sn-bibliography}
	
\end{document}